 \documentclass{aa}
\usepackage{graphicx}
\usepackage{txfonts}
\newcommand{\sax}{{\it BeppoSAX}}
\newcommand{\suza}{{\it Suzaku}}
\newcommand{\sw}{{\it Swift}}
\newcommand{\glast}{{\it GLAST}}
\newcommand{\agl}{{\it AGILE}}
\newcommand{\inte}{{\it INTEGRAL}}
\newcommand{\Ch}{{\it Chandra}}
\newcommand{\XN}{{\it XMM-Newton}}

\newcommand{\xr}{X-ray~}

\begin{document}

\title{The complete catalogue of GRBs observed by the Wide Field Cameras 
on board BeppoSAX}

\author{L. Vetere\inst{1,2}
\and P. Sof\mbox{}f\mbox{}itta\inst{3}
\and E. Massaro\inst{4,1} 
\and P. Giommi\inst{1} 
\and E. Costa\inst{3} 
}
	
\institute{
ASI Science Data Center, ASDC c/o ESRIN, via G. Galilei, I-00044 Frascati, Italy 
\and
{\it current:} Department of Astronomy and Astrophysics, Pennsylvania State University, 
5252 Davey laboratory, University Park, PA 16802 USA 
\and 
INAF-IASF, Sezione di Roma, Via del Fosso del Cavaliere, I-00133, Roma, Italy
\and
Dipartimento di Fisica, Sapienza Universit\`a di Roma, Piazzale A. Moro 2, I-00185, 
Roma, Italy}

\offprints{~~~~~~~~\\
P.~Giommi: ~paolo.giommi@asdc.asi.it}

\date{Received:.; accepted:.}

\titlerunning{The complete GRB catalogue observed by WFC on board BeppoSAX}
\authorrunning{L. Vetere et al.}

\abstract{ We present the complete on-line catalogue of gamma-ray bursts observed by the two Wide Field Cameras on board \sax~ in the period 1996--2002.}{ Our aim is to provide the community with the largest published data set of GRB's prompt emission \xr light curves and other useful data.}
{ This catalogue (BS-GRBWFCcat) contains data on 77 bursts and a collection of the \xr light curves of 56 GRB discovered or noticed shortly after the event and of other additional bursts detected in subsequent searches.
Light curves are given in the three \xr energy bands (2--5, 5--10, 10--26 keV).}{ The catalogue can be accessed from the home web page of the ASI Science Data Center-ASDC ({\it http://www.asdc.asi.it}).}{}


\maketitle


\section{Introduction}
The \sax~ satellite, an Italian mission for X-ray astronomy with a contribution 
from the Netherlands Agency for Aerospace Programs (NIVR) and the Space Research 
Organisation Netherlands (SRON) (Boella et al. 1996), was launched on 
April 30, 1996 and remained operative for six years up to April 30, 2002.
The scientif\mbox{}ic payload consisted of two Wide Field Instruments (WFI) and four 
Narrow Field Instruments (NFI).
The WFI instruments consisted of two Wide Field Cameras (WFC), pointing at opposite 
directions in the sky and perpendicular to the NFIs, and the Gamma-ray Burst 
Monitor (GRBM).

One of the most important results obtained by the \sax~ mission was the discovery 
of the X-ray afterglows of GRBs (Costa et al. 1997) which led to the detection of the associated optical emission (van Paradijs et al. 1977) and to the measurement 
of GRB redshifts (Djorgovski et al. 1997). 
In the six years of the \sax~ lifetime the WFIs detected several GRBs and provided 
light curves down to the energy of 2 keV.
This also led to the discovery of a new class of bursts, called X-ray Flashes (XRF), 
because these are much brighter at energies lower than 30 keV than in the low energy 
gamma rays (Heise et al. 2001).

Up to now this is the largest published data set of GRB \xr light curves  useful 
to investigate their time and spectral properties, particularly in comparison with 
data at higher energies.
We have collected all these data in a catalogue, named BeppoSAX GRB-WFC Catalogue 
(shortly BS-GRBWFCcat), available on-line either from the Home page of the ASDC (ASI 
Science Data Center) at the URL: {\it http://www.asdc.asi.it} or at the direct URL
{\it http://www.asdc.asi.it/grb$_{-}$wfc}.
This catalogue is complementary to that of GRBM presented by Guidorzi et 
al. (2004), which covers the energy range 40-700 keV.
In this note we give a brief presentation of the BS-GRBWFCcat and of its content.
More information on the use of this catalogue is available directly from the web 
pages given above.

\section{Wide Field Cameras}

Each WFC (Jager et al. 1997) was a Multi-Wire Proportional Counter with an open area 
of 25$\times$25 cm$^{2}$ and a useful energy range of 2--26 keV.
The two-dimensional position- and energy-sensitive counter was placed behind an opaque 
screen with a pseudo random array of holes, a so-called shadow or coded mask. 
X-ray sources in the f\mbox{}ield of view cast shadows of the mask on the detector, each
displaced according to the position of the source. 
A cross-correlation (Fenimore et al. 1978) carried out on the ground, reconstructs the 
positions and f\mbox{}luxes of all sources in the f\mbox{}ield. 
The WFCs had a good sensitivity over a large f\mbox{}ield of view and a good angular resolution 
within the entire f\mbox{}ield (due to small mask holes and to the good resolution of the detector).
The transparency of the mask was 33\% and the f\mbox{}ield of view was limited by a collimator 
with no f\mbox{}lat f\mbox{}ield to 40$^{\circ} \times$ 40$^{\circ}$. 
The ef\mbox{}fective area was $\sim$150 cm$^{2}$ at the centre of the f\mbox{}ield of view.
The sensitivity for burst detection has been estimated at a f\mbox{}lux level of
$\sim$4$\times$10$^{-9}$ erg cm$^{-2}$ s$^{-1}$ (De Pasquale et al. 2006).

\section{Observations and data reduction}
All the GRB light curves (LC) in the BS-GRBWFCcat were taken with the two WFCs on board 
\sax.  
There are two dif\mbox{}ferent ways to extract a LC from the WFC data:
\begin{itemize}
\item  $i)$ consider the whole detector image and subtract the background component 
deconvolving it using the mask response function;
\item $ii)$ select only the region concerned by the burst, without subtraction 
of the background component, including the pixel occulted by the mask in the 
direction of the source.
\end{itemize}

The f\mbox{}irst method gives LCs without background, while the second method gives LCs where counts are distributed according to the Poisson statistics. 
Note that in the majority of WFC f\mbox{}ields the integrated count rate of instrumental background 
is $\sim$130 counts/s 85\% of which are due to the cosmic  X-ray background and undetected sources.
In a few sky areas, like the Galactic Centre or the Cygnus region, this value is signif\mbox{}icantly 
higher and can even double.
The burst's count rate depends not only on the incident f\mbox{}lux and spectrum but also on the 
mask transparency, the detector ef\mbox{}f\mbox{}iciency and on the burst location in the f\mbox{}ield of view.
We decided, therefore, to use the f\mbox{}irst method of LC extraction which gives a  
controlled background subtraction.
For each burst we created three LCs in the three energy bands 2--5, 5--10, 10--26 keV, 
selected to have approximately the same number of counts in each band for a typical GRB. 
The integration time was usually 1.0 s.

\section{Catalogue description}
GRB data in the BS-GRBWFCcat are given in two tables.
The f\mbox{}irst one contains general information and gives access to LCs of 62 bursts (two are
classif\mbox{}ied as X-ray Flashes, XRF) observed from July 1997 to February 2002: 
56 of these GRBs were either triggered by the onboard software of the GRBM or discovered through the quick look analysis or by alerts from other satellites, while the other 6 were detected in a subsequent analysis of GRBM count rates.
The latter bursts are indicated as {\it of\mbox{}f-line} events and are marked  
by an asterisk. No information on them can be found in the GCN archive and no 
follow-up observations were performed.

The second table contains 15 bursts found in the of\mbox{}f-line analysis for which only very 
few data are available in the literature (Kippen et al. 2004 (KP04); in't Zand et al. 
2004 (TZ04); Heise \& in't Zand 2004 (HTZ04); D'Alessio, Piro \& Rossi 2006 (DAPR), 
Rossi et al. 2005 (RS05)) and many of them are classif\mbox{}ied as XRF (reference codes in 
parentheses are those reported on the web page).  
DAPR, in particular, published a catalogue of XRF-XRR bursts selected according to the 
criterion proposed by Lamb \& Graziani (2003) where the discriminant parameter is the 
hardness ratio $H_h$ between the f\mbox{}luence in the 2--30 keV energy range and that measured 
in the 30--400 keV range. 
When $H_h$ is in the interval 0.32--1.0 the burst is classif\mbox{}ied as an X-Ray Rich GRB (XRR),
while for $H_h$ higher than 1.0 it is an XRF.
The classif\mbox{}ication in the DAPR catalogue is also given.

All GRBs are listed in chronological order and are named in the usual way,
i.e. GRB or XRF followed by the date.
For the GRBs in the f\mbox{}irst table the following information is given:
trigger MJD and UT time, duration, coordinates (J2000), peak f\mbox{}lux in the 2--10 keV range 
(from Frontera 2004), the redshift, when known, with the relative reference, and some short notes.

  \begin{figure}
 \begin{center}
\hspace{0.cm}
    \includegraphics[width=0.27\textwidth,angle=-90]{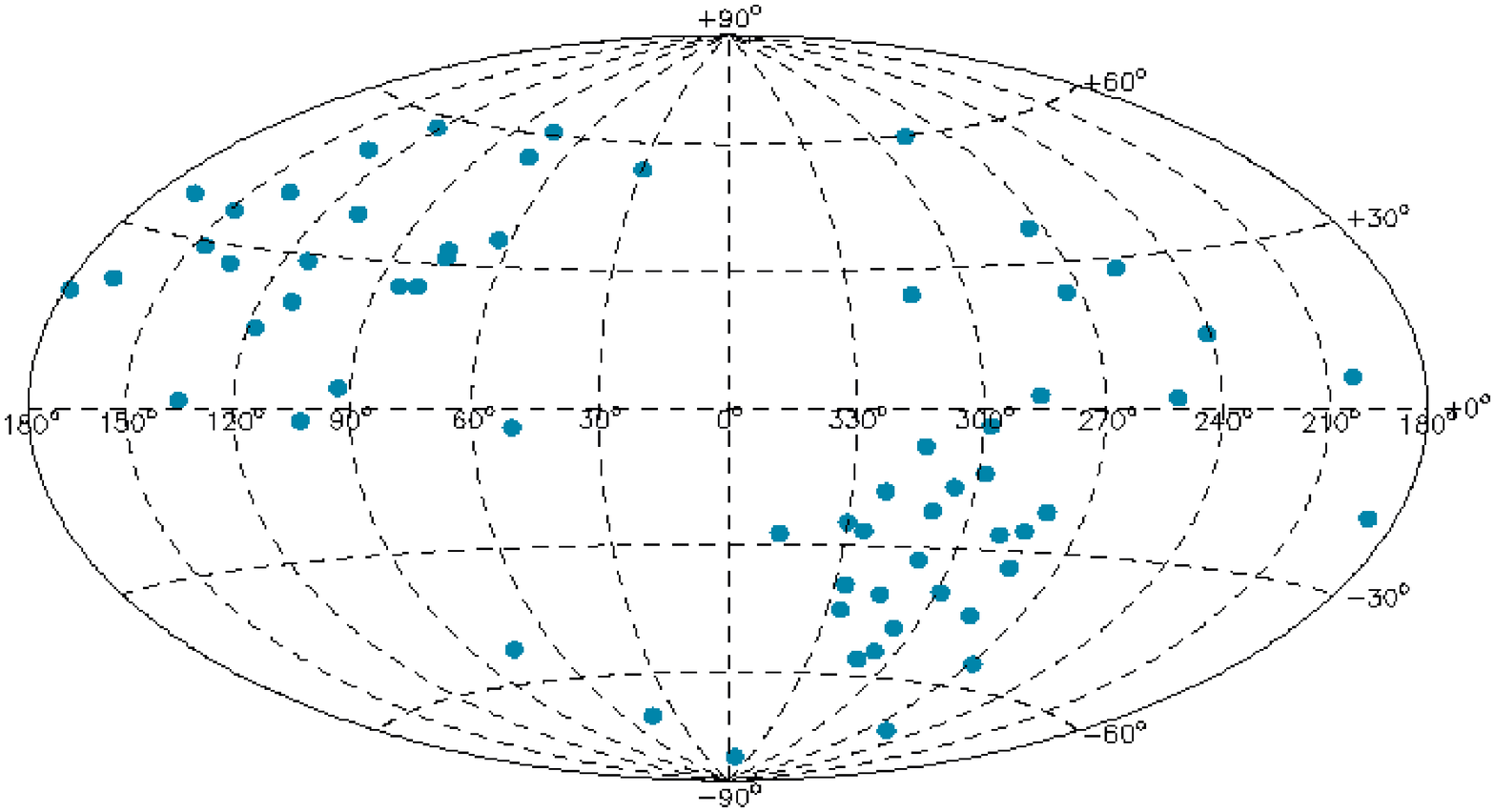}
    \caption{\label{fig1} The sky distribution in galactic coordinates of the GRBs
     in the Section 1 of BS-GRBWFCcat.
     }
 \end{center}
 \end{figure}

  \begin{figure}
 \begin{center}
\vspace{-2.cm}
 \hspace{-1.cm}
   \includegraphics[width=0.4\textwidth]{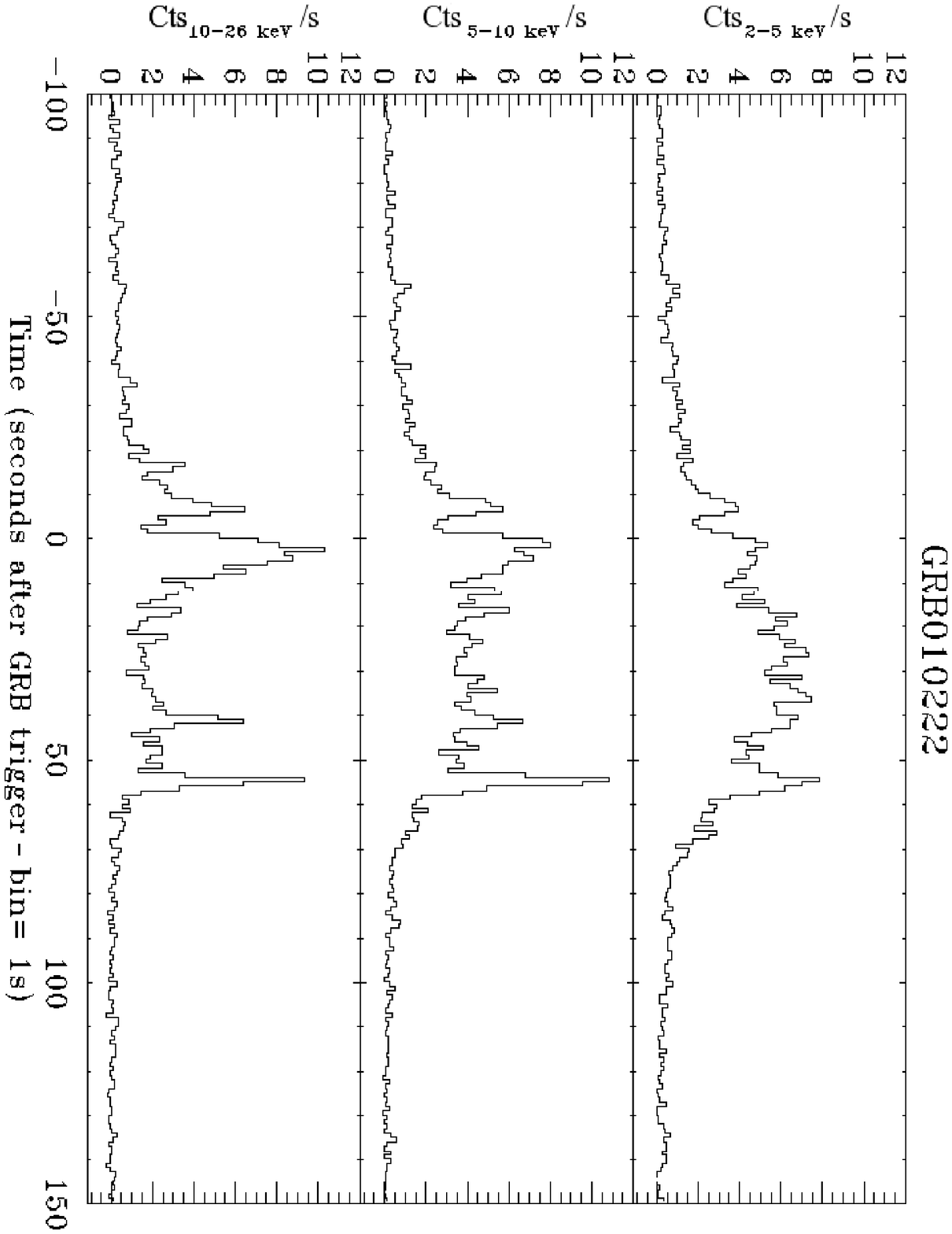}
    \caption{\label{fig2} An example of the three X-ray light curves given in the three 
     energy ranges (2--5, 5--10, 10--26 keV top to bottom) in the BS-GRBWFCcat.
     Here are shown the LCs of GRB~010222, characterized by a time evolution strongly
     dependent on energy. 
     Note the signif\mbox{}icant excess of counts before the trigger time. }
 \end{center}
 \end{figure}

The sky distribution in galactic coordinates of all GRBs with known coordinates listed 
in the BS-GRBWFCcat is shown in Fig. 1:
it is highly anisotropic with two major concentrations in the direction of the ecliptic poles.
This is due to the non uniform exposure of WFC that pointed in a direction perpendicular
to NFIs.
The same f\mbox{}igure also appears on the web page of Table 1 as an interactive slot.
By clicking on the plot one can open a window and on the individual GRB positions it is 
possible to see the LCs. 

It is possible to select a GRB in the f\mbox{}irst table of BS-GRBWFCcat and to visualize three background subtracted WFC light curves (1 s time bin) in the energy bands given above.
It is also possible to retrieve the LC data as ascii f\mbox{}iles. 
In the case of GRB000214, however, the only available LC is in entire energy range 2-26 keV 
without background subtraction.
The zero time of each light curve was f\mbox{}ixed at the trigger, but sometimes a clear signal is 
detectable well before this time and, for this reason, LCs start several tens of seconds before 
the trigger.
Only the LC of the famous GRB 990123 is incomplete because its radiation was strongly 
absorbed by the Earth atmosphere at the end of the LC.
An example of a GRB light curve is shown in Fig. 2.

The durations of GRBs given in the catalogue are not the $T_{90}$ frequently reported in the 
literature.
We estimated the duration of the 2-5 keV prompt emission from the time integrated (cumulative) light curve, considering the duration of the interval where it differs from a constant value by means of linear fits. 
These estimates, however, depend on the S/N ratio and are uncertain in some cases. 
Moreover, in several events there is a clear indication of a pre/post burst emission and the beginning and the end of the burst cannot be clearly distinguished from background 
f\mbox{}luctuations.
In any case, the durations must be considered as an estimate of the time in which we
have a detectable signal in the X-ray LC of a GRB.
Only for two bursts (GRB 010220 and GRB 02032,1 which have LCs highly changing with
energy) the durations were estimated in the 10-26 keV range because the signal in the 
lowest band was too low for a reliable measure.
Some short comments on the duration estimates are also given in the table.
The shortest event is about 13 s long, whereas the longest one reaches 248 s.
A histogram of the burst durations in the catalogue is shown in Fig. 3.
More precise evaluations can be derived by the users from the LC data using 
criteria consistent with the particular problem under investigation.  

   \begin{figure}
 \begin{center}
\hspace{0.cm}
    \includegraphics[width=0.42\textwidth,angle=-90]{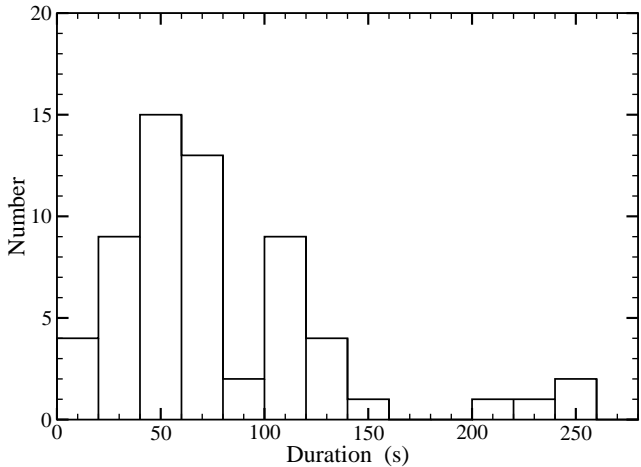}
    \caption{\label{fig3} Histogram of the GRB duration in the BS-GRBWFCcat
             estimated from the 2--10 keV light curves.}
 \end{center}
 \end{figure}

For the 10 GRBs studied by Vetere et al. (2006) we reported also the X-ray photon indices derived from the hardness ratios between the three energy bands used.
We plan to add this information, at least for the bursts with a good S/N ratio, in future updates of the catalogue.   

The BS-GRBWFCcat also provides a link to the GRB web page at MPE-Garching (care of  
J. Greiner) when available. 
Tools for an on-line spectral analysis are planned
to be available in future versions of the BS-GRBWFCcat. 

\section{Conclusion}
The nature of sudden explosions producing GRBs is one of the most mysterious phenomena in the universe.
Despite several space instruments have collected up to now a large amount of data, a full understanding of the physical processes occurring in these huge explosions has not been reached yet.
To provide the astrophysical community with some useful data on these events we prepared a web accessible catalogue (BS-GRBWFCcat) containing the X-ray LCs of all GRBs detected by the Wide Field Cameras on board \sax.
This table includes information on several events, mainly of XRF type, discovered through an of\mbox{}f-line analysis.
Unlike other catalogues, it provides data in the 2-26 keV band that are generally not 
available for the majority of GRBs.
It is remarkable that X-ray LCs are generally dif\mbox{}ferent from those at higher energies suggesting 
that either other emission components or physical processes are relevant in this band. 
 
Our aim is to extend the observational knowledge of this phenomenon while
a mission like \sw, mainly devoted to GRB studies, and others such as \inte, \suza, \Ch~ and \XN~ are currently providing new important data. Even more data are expected in the near future when the $\gamma$-ray window will be explored by \agl~ and \glast.

\begin{acknowledgements}
This work has been partially supported by INAF under the ASI contract for the
scientif\mbox{}ic activity at ASDC.
\end{acknowledgements}

\end{document}